\documentclass{article}

\PassOptionsToPackage{numbers, compress}{natbib}

\usepackage[preprint]{neurips_2025}

\usepackage[utf8]{inputenc} 
\usepackage[T1]{fontenc}    
\usepackage{hyperref}       
\usepackage{url}            
\usepackage{booktabs}       
\usepackage{amsfonts}       
\usepackage{nicefrac}       
\usepackage{microtype}      
\usepackage{xcolor}         

\usepackage{soul,framed} 
\usepackage{multicol}
\usepackage{multirow}
\colorlet{shadecolor}{yellow}
\usepackage[pdftex]{graphicx}
\graphicspath{{../pdf/}{../jpeg/}}
\DeclareGraphicsExtensions{.pdf,.jpeg,.png}
\usepackage[cmex10]{amsmath}
\usepackage{array}
\usepackage{mdwmath}
\usepackage{lscape}
\usepackage{mdwtab}
\usepackage{eqparbox}
\usepackage{float}
\usepackage{bm}
\usepackage{lineno}
\usepackage{makecell}
\usepackage{tikz}
\usetikzlibrary{positioning}
\usepackage{calc}
\usepackage{diagbox}
\usepackage{threeparttable}
\usepackage{verbatim}
\usepackage{appendix}
\usepackage[figuresright]{rotating}
\usepackage{natbib}
\usepackage{todonotes}
\usepackage{xspace}
\usepackage{wrapfig}
\newcommand{\myname}{CAW\xspace}
\newcommand{\Cname}{contextual generation states-aware watermark capacity evaluator\xspace}
\newcommand{\xhdr}[1]{{\noindent\bfseries #1}.}

\title{Enhancing Watermarking Quality for LLMs via Contextual Generation States Awareness}

\author{%
Peiru Yang\thanks{Tsinghua University. Email: \texttt{ypr21@mails.tsinghua.edu.cn}} \quad
Xintian Li\footnotemark[1] \quad
Wanchun Ni\thanks{Apple} \quad
Jinhua Yin\footnotemark[1] \quad
Huili Wang\footnotemark[1] \\
Guoshun Nan\thanks{Beijing University of Posts and Telecommunications} \quad
Shangguang Wang\footnotemark[3] \quad
Yongfeng Huang\footnotemark[1] \quad
Tao Qi\thanks{Beijing University of Posts and Telecommunications. Corresponding author. Email: \texttt{taoqi.qt@gmail.com}}
}

\begin{document}

\maketitle

\begin{abstract}

Recent advancements in watermarking techniques have enabled the embedding of secret messages into AI-generated text (AIGT), serving as an important mechanism for AIGT detection. 
Existing methods typically interfere with the generation processes of large language models (LLMs) to embed signals within the generated text. 
However, these methods often rely on heuristic rules, which can result in suboptimal token selection and a subsequent decline in the quality of the generated content.
In this paper, we introduce a plug-and-play contextual generation states-aware watermarking framework (\myname) that dynamically adjusts the embedding process. 
It can be seamlessly integrated with various existing watermarking methods to enhance generation quality.
First, \myname incorporates a watermarking capacity evaluator, which can assess the impact of embedding messages at different token positions by analyzing the contextual generation states. 
Furthermore, we introduce a multi-branch pre-generation mechanism to avoid the latency caused by the proposed watermarking strategy.
Building on this, \myname can dynamically adjust the watermarking process based on the evaluated watermark capacity of each token, thereby minimizing potential degradation in content quality.
Extensive experiments conducted on datasets across multiple domains have verified the effectiveness of our method, demonstrating superior performance compared to various baselines in terms of both detection rate and generation quality.

\end{abstract}


\section{Introduction}\label{section1}

Watermarking has become a crucial method for detecting AI-generated text (AIGT), addressing growing concerns over LLM misuse and content authenticity~\cite{perkins2023academic, ren2024copyright, grinbaum2022ethical}.
Mainstream watermarking methods embed perturbations as identification tags into LLM outputs through modifications to the models' generation mechanisms~\cite{wang2024building, kirchenbauer2023watermark, aronson2022watermarking, zhao2023provable}.
Existing LLMs adhere to a similar framework, wherein the model first computes the probability distribution over the vocabulary, and then selects the next token based on certain sampling strategies.
To incorporate secret messages into generated content, watermarking methods generally rely on heuristic strategies to reweight the probability distribution or modify the sampling strategy.
For example, ~\citet{kirchenbauer2023watermark} propose a watermark that modifies token probability by dividing the vocabulary into ``red'' and ``green'' tokens, and promote the use of green tokens during sampling.
However, the perturbation introduced by the watermark methods can result in the sub-optimal selection of tokens and even lead to serious errors in the generated content, which can substantially degrade the generation quality.
Such impacts on text quality may diminish the willingness of model owners and publishers to deploy watermarks.

\begin{figure}
    \centering
    \includegraphics[width=\linewidth]{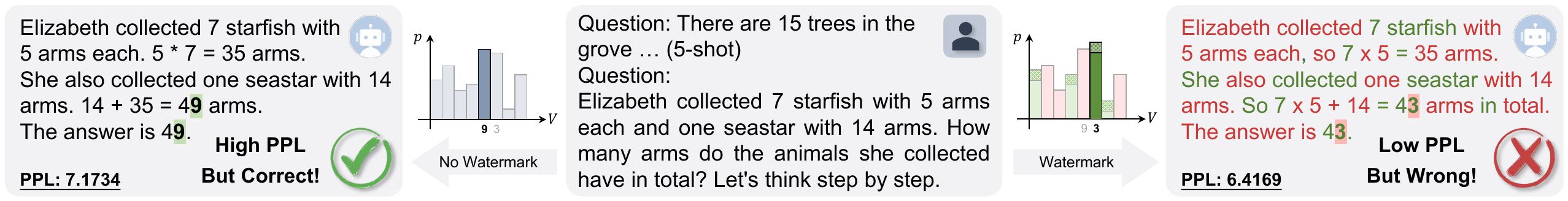} %
    \vspace{-5pt}
    \caption{Comparison between watermarked and unwatermarked outputs on Llama2-13b with GSM8K dataset.
    Watermarking introduces sampling disturbances that distort key information and cause incorrect answers, which the perplexity metric fails to capture.}
    \label{fig:intro}
\end{figure}

Although existing methods have explored preserving output quality, they mostly rely on isolated token confidence, assuming high-confidence tokens are irreplaceable~\cite{kirchenbauer2023watermark, lee2023wrote, lu2024entropy, wouters2024optimizing, liu2024adaptive}.
Based on this, tokens with high-confidence distributions are typically exempt from watermarking.
However, this mechanism relies solely on isolated token confidence, which may misidentify irreplaceable generated content due to localized informational bias or insufficiency.
In particular, complex tasks near or beyond LLM knowledge boundaries may result in low confidence for critical tokens.
Thus, these methods still encounter watermark-induced quality degradation.
For example, Fig~\ref{fig:intro} shows the responses of the Llama2-13b model to a math problem from the GSM8K dataset with and without KGW watermarking\cite{kirchenbauer2023watermark}.
The model originally generates a correct answer, yet watermarking leads to suboptimal sampling at key positions, causing computational errors.
While red-highlighted tokens denote unwatermarked high-confidence tokens, their confidence levels alone cannot reliably identify critical information to ensure output quality, as clearly demonstrated in Fig~\ref{fig:intro}.
Thus, further research is needed to enhance the effectiveness of watermarking output quality.

Furthermore, while some studies claim that their watermarking methods are quality-preserving based on perplexity evaluations\cite{kirchenbauer2023watermark, hu2023unbiased, wu2024resilient}, our findings indicate that such assessments are inadequate for capturing fine-grained flaws.
Perplexity reflects a model's predictive ability by averaging inverse probabilities over the output.
However, the actual quality of the output may be dominated by a small number of critical tokens.
Specifically, errors in these pivotal elements can severely degrade the entire text's quality, while perplexity may not vary substantially since the fluency and coherence levels remain unchanged.
For instance, as shown in Fig.~\ref{fig:intro}, the watermarked output contains errors in critical figures that result in an incorrect overall response. 
Nevertheless, the response remains coherent, resulting in no significant difference in its perplexity metric.
In this case, perplexity cannot indicate the accuracy of generated content and thus fails to measure text quality comprehensively.
In conclusion, the selection of an appropriate evaluation protocol is crucial for the design of quality-preserving watermarking techniques aimed at enhancing the accuracy of generated content.

To address the challenges above, we propose a plug-and-play watermarking framework with contextual generation states awareness (named \myname), which can be seamlessly integrated with various existing watermarking techniques to improve text quality.
The core idea of our work is to adaptively adjust the watermarking embedding process through awareness of the impact on the generation quality.
First, \myname incorporates a watermark capacity evaluator to assess the impact of embedding messages at different token positions through contextual generation states awareness.
Specifically, we define watermark capacity as the importance level of semantic information carried by a token, which equivalently indicates its bearable watermark strength.
Preliminary experimental results show that contextual generation states are important for assessing watermark capacity, and the synergistic incorporation of the current token and its contextual generation states yields optimal model performance.
In addition, a multi-branch pre-generation mechanism is employed in \myname to mitigate the impact of combining both past and future generation states on generation latency.
This mechanism utilizes tree attention\cite{miao2024specinfer} to pre-generate potential sampling strategies and their corresponding probability branches, enabling efficient decoding iterations.
As a plug-and-play framework, \myname can be seamlessly integrated with existing watermarking methods to minimize potential quality degradation without compromising their original structure, thus showing good generalization ability and high practicality.
We conduct extensive experiments across two real-world datasets and four foundation models to evaluate the detection rate and text quality of \myname.
Experimental results demonstrate that, compared to SOTA baselines, \myname achieves superior performance in terms of both watermarking detection rate and generation accuracy, while maintaining the watermarking robustness and efficiency.

\section{Related Works}\label{section2}

\xhdr{Reweighting-based Watermarking}
Reweighting-based watermarks work by modifying the model's output logits or probability distribution during inference before sampling.
\citet{kirchenbauer2023watermark} propose the first LLM watermarking method using logits reweighting, which randomly splits the vocabulary into green and red lists, increasing the selection probability of green tokens.
Later works improve this approach from different angles~\cite{liu2023unforgeable, zhao2023provable}.
\citet{liu2023unforgeable} uses a separate neural network for vocabulary partition to enhance unforgeability.
\citet{zhao2023provable} adopts a fixed partition instead of hashing previous tokens, improving security guarantees.
Some studies extend watermarking to multi-bit formats for embedding more information~\cite{wangtowards, yoo2024advancing, guan2024codeip}.
Others focus on improving robustness against attacks like paraphrasing~\cite{ren2023robust, zhang2024remark}.
These studies primarily emphasize method robustness, typically evaluating text quality using only basic perplexity measurements. 
Consequently, achieving an optimal balance between text quality and robustness remains challenging.

\xhdr{Sampling-based Watermarking}
Unlike reweighting-based methods, sampling-based watermarking preserves the original token probability distribution and instead modifies the sampling procedure.
\citet{aronson2022watermarking} propose the first such method using a pseudo-random hash to alter token sampling.
\citet{kuditipudi2023robust} introduce two distortion-free sampling-based watermarking strategies.
\citet{christ2024undetectable} present cryptographically-inspired undetectable watermarks using a pseudo-random function.
\citet{fu2024gumbelsoft} extend \citet{aronson2022watermarking} by introducing uncertainty to enhance output diversity.
\citet{dathathri2024scalable} propose tournament sampling and validate it on commercial LLMs.
\citet{zhu2024duwak} design a dual-watermarking method using contrastive search.
These methods also face the common challenge of balancing quality and robustness.

\xhdr{Existing Efforts on Enhancements of Text Quality}
Recent efforts have aimed to improve text quality during watermarking, using methods such as entropy-based token selection~\cite{lee2023wrote,lu2024entropy,liu2024adaptive}, and semantic-aware vocabulary partitioning~\cite{fu2024watermarking, chen2024watme}.
However, entropy-based methods may misidentify key tokens because they evaluate each token in isolation, ignoring contextual semantics that determine token importance in meaning.
Meanwhile, vocabulary-partitioning strategies are inherently tied to reweighting-based mechanisms and thus cannot be applied to sampling-based watermarking, limiting their generalizability across different frameworks.
Overall, most techniques use heuristic token manipulation, often degrading semantic quality.
We instead explore contextual generation states for more adaptive, quality-preserving watermarking (see Appendix~\ref{RL} for details).

\section{Method}\label{section3}
\subsection{Framework Overview} \label{subsection32}
\begin{figure*}[t]
    \centering
    \includegraphics[width=\textwidth]{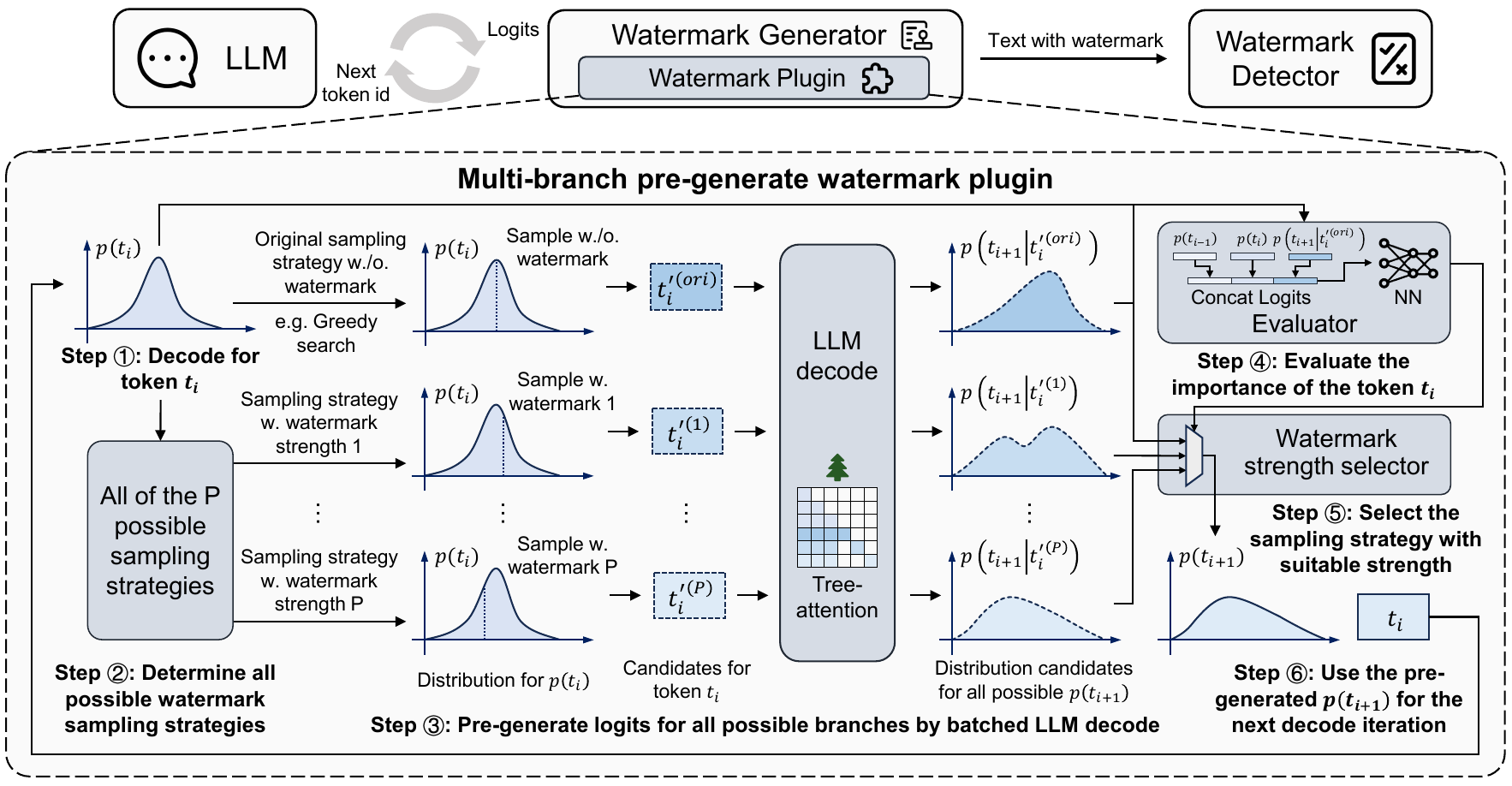} 
    \vspace{-10pt}
    \caption{Architectural overview of \myname, highlighting its plug-and-play integration capability.
    \myname dynamically optimizes watermark embedding through watermark capacity evaluation with contextual generation states awareness and the multi-branch pre-generation mechanism.
}
    \label{fig:method}
\end{figure*}

This section presents the framework of \myname step by step, illustrating its adaptive watermarking process (Fig~\ref{fig:method}).
During LLM decoding, \myname operates as a plugin that determines whether and how strongly to inject a watermark at each token position.
This decision is guided by the token’s watermark capacity, defined as its semantic importance and tolerance to perturbation.
Hence, \myname incorporates a watermark capacity evaluator based on contextual generation states, namely the probability distribution of previous and subsequent token positions, to assess the impact of watermark injection at each token position.
The watermark capacity prediction is then used to determine whether to inject a watermark and the intensity of the watermark. 
To reduce latency from using future context, a multi-branch pre-generation mechanism is adopted.
Tree attention enables efficient pre-computation of logits for multiple candidates, improving both inference speed and memory usage.

Here, we will gradually introduce the overall procedure in 6 steps:
\textbf{Step 1}: Decode $\mathcal{M}$ at position $i$ to obtain the token distribution $p(t_i) \in \Delta(\mathcal{V})$.
\textbf{Step 2}: Determine all $P$ potential watermark sampling or reweighting strategies at position \( i \), where \(P\) represents the total number of possible strategies.
The strategy here is determined by the variable controlling watermark strength in various methods, with different strategies representing different watermarking strengths.
We define the sequence of all candidate tokens at position \( i \) as \( \mathbf{t_i^{\prime}} = [t_i^{\prime(ori)}, t_i^{\prime(1)}, ..., t_i^{\prime(P)}]\),  
For each strategy, we get the corresponding candidate for position \( i \) to form the candidate token sequence \(\mathbf{t_i^{\prime}}\).
\textbf{Step 3}: For each candidate token in $\mathbf{t_i^{\prime}}$, pre-generate logits at position $i+1$.
Perform tree attention decoding \cite{miao2024specinfer} to obtain all candidate distributions \(p(t_{i+1})\), namely $[p(t_{i+1}\mid t_i^{\prime(ori)}), p(t_{i+1}\mid t_i^{\prime(1)}), ..., p(t_{i+1}\mid t_i^{\prime(P)})]$,
where \( p(t_{i+1} \mid \cdot) \) represents the probability distribution at position \( i+1 \) given the token determined at position \( i \).
\textbf{Step 4}: Combine the distribution at position \(i\) with the distributions of its preceding and subsequent contextual positions.
Input the joint distribution \([p(t_{i-1}), p(t_i), p(t_{i+1} \mid t_i^{\prime(ori)})]\) into the watermark capacity evaluator $\mathcal{E}$ for $\hat{C_i}$, namely the contextual generation states aware watermark capacity at position $i$.
\textbf{Step 5}: Based on the predicted watermark capacity \(\hat{C_i}\), select the appropriate watermark strength and further determine the corresponding watermarking strategy from the $P$ candidates in Step 2.
The token \( t_i \in \mathbf{t_i^{\prime}} \) is finalized.
\textbf{Step 6}: Use the pre-generated \(p(t_{i+1} \mid t_i)\) to proceed to the next decoding iteration.
In this way, \(p(t_{i+1})\) has been pre-generated, so it will not cause additional latency.
The six key steps outlined above form a complete decode iteration for \myname as shown in Fig~\ref{fig:method}, which can be seamlessly and efficiently integrated with both reweighting-based and sampling-based watermarking methods.

\subsection{Watermark Capacity Evaluator} \label{subsection33}

\xhdr{Motivation and Design}
This subsection introduces the contextual generation states-aware watermark capacity evaluator, a core component of the whole framework.
It evaluates contextual generation states to estimate the impact of watermark embedding at each token, enabling \myname to adaptively watermark with minimal quality degradation.

We define watermark capacity as the significance of the semantic information conveyed by a token, reflecting its tolerance to watermark strength and inversely related to embedding impact.
It is inversely related to the impact of embedding secret information at the token position. 
Intuitively, this capacity depends on both the token and its surrounding context.
Accordingly, we design a watermark capacity evaluator that estimates embedding impact at each position by analyzing contextual generation states in real time.
Figure~\ref{fig:classifierStructure} in Appendix~\ref{WCE} illustrates the structure of the proposed watermark capacity evaluator.
We use the contextual probability distribution during generation as input and output the watermark capacity at token position $t_i$.
Letting the evaluator be $\mathcal{E}$, the process is:
\begin{equation}
\label{eq:evaluator}
\hat{C_i} = \mathcal{E}\left( \left[ p(t_{i+k}) \right]_{k=-N^-}^{N^+}; \Theta^{*} \right),
\end{equation}
where $\hat{C_i} \in (0,1)$ is the predicted capacity, and $\Theta^{*}$ is the trained parameter.
$-N^-, N^+$ determine the contextual range in input.
We implement $\mathcal{E}$ as a 3-layer fully connected network.
To train and evaluate it, we define a task using human-written responses as references, with GPT-4o marking quality-critical segments as ground truth.
We denote $\left[ p(t_{i+k}) \right]_{k=-N^-}^{N^+} $ as $\mathbf{p}(t_i)$ for brevity.

\begin{figure*}[t]
    \centering
    \includegraphics[width=\textwidth]{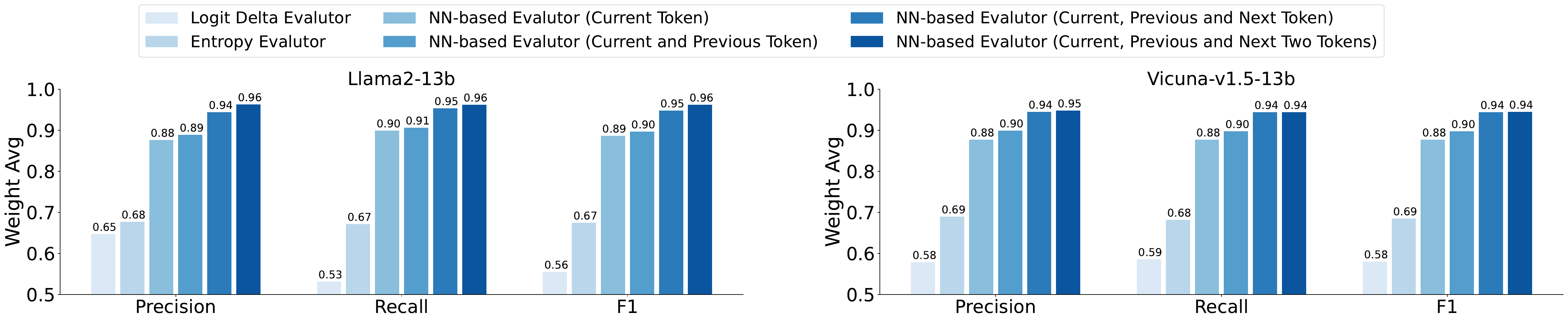} %
    \vspace{-10pt}
    \caption{
    Comparison of watermark capacity modeling methods (Left: Llama2-13B; Right: Vicuna-13B), showing that contextual states awareness greatly improves performance.
    }
    \label{fig:ablation}
\end{figure*}

\xhdr{Preliminary Experiments}
To determine the optimal evaluator, we conduct preliminary experiments for comparative analysis. 
These experiments investigate different evaluators and the influence of context length on performance.
For baseline comparison, we employ the logit delta evaluator and the entropy evaluator. 
All evaluators utilize the probability distribution generated by the LLM to predict watermark capacity at each token position.
Specifically, the logit delta evaluator measures the gap between the top-1 and top-2 logits; a larger delta suggests stronger model preference and greater impact from watermarking
This approach is similar to the intrinsic evaluation mechanisms found in vocabulary-based methods, such as the KGW approach \cite{kirchenbauer2023watermark}. 
The entropy evaluator computes token-level entropy, where lower values indicate higher determinism and thus lower watermark capacity~\cite{lee2023wrote, lu2024entropy, liu2024adaptive}.
The structure of our proposed evaluator has been previously described.
Due to the typically large vocabulary size of LLMs, we only utilize a limited number of values (100 in our experiments) from the probability distribution, specifically those corresponding to the highest logits at a token position, as input features in our analysis.

Fig~\ref{fig:ablation} presents performance comparisons of various watermark capacity evaluators using precision, recall, and F1 metrics.
Experimental analysis yields three key findings.
First, the neural network-based evaluator outperforms traditional logit delta and entropy methods, benefiting from its ability to process multidimensional information.
Unlike manually crafted statistical features, it uses raw ranked probability distributions and learns patterns via supervised learning.
Second, adding contextual information significantly improves accuracy over isolated-token input, confirming the value of contextual generation states for capacity estimation.
Third, we explore how context window size affects performance.
Integrating contextual probabilities is consistently beneficial, with longer historical and future contexts (larger $N^-$ and $N^+$) yielding gradual accuracy gains.
However, expanding the window also increases latency due to future-token dependencies in causal decoding.
The best trade-off is achieved with immediate left and right context ($N^- = 1$, $N^+ = 1$), as further expansion offers limited returns.
Overall, neural networks effectively leverage contextual generation states to outperform statistical baselines, striking a balance between informativeness and efficiency through targeted context window design.
This approach provides stronger modeling of token-level watermark capacity than manual feature-based methods.

\xhdr{Contextual Generation States-Aware Watermarking}
\myname adaptively adjusts watermark strength based on the predicted token capacity. Tokens with $\hat{C_i} > \theta$ are left unmodified, preserving the original probability distribution and sampling strategy to protect semantically important content. For tokens with $\hat{C_i} < \theta$, watermarking is applied with strength scaled proportionally to capacity. 
This adaptive mechanism can be integrated with different watermarking schemes: for sampling-based methods, the top-K set is dynamically narrowed based on $\hat{C_i}$; for reweighting-based methods, the bias $\delta$ applied to token logits is scaled accordingly. The specific integration strategies with both categories are detailed in Appendix~\ref{CGSA}.

\subsection{Multi-branch Pre-generation Mechanism}
\begin{wrapfigure}{htbpr}{0.5\linewidth} 
    \centering
    \includegraphics[width=1\linewidth]{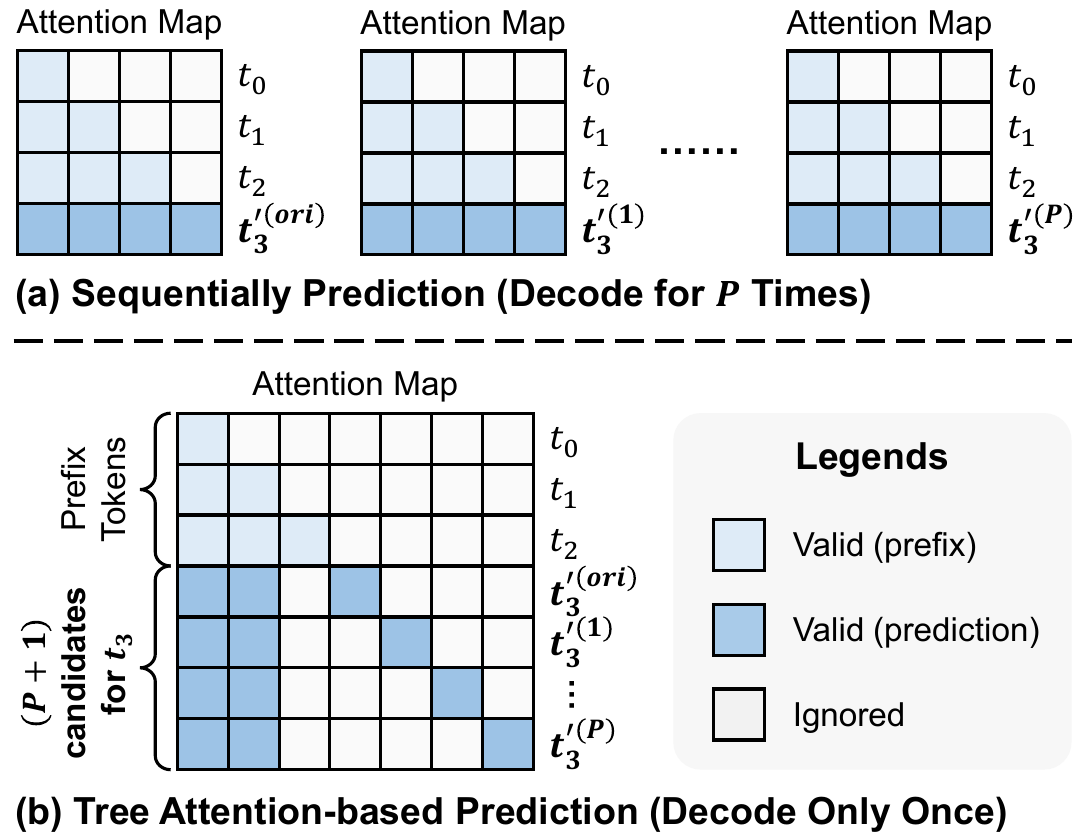} 
    \vspace{-10pt}%
    \caption{    
    Details of tree attention decoding. (a) Sequential attention for each $t_3$ in $\textit{t}_3^{\prime}$ requires $P+1$ computations. (b) Tree attention mask enables a single computation for all candidates by zeroing others’ mask values.
}
    \label{fig:tree-attn}
\end{wrapfigure}

To mitigate the latency introduced by evaluating future token distributions in contextual watermark capacity estimation, \myname incorporates a multi-branch pre-generation mechanism based on tree attention, as illustrated in Fig~\ref{fig:tree-attn}. At each decoding step, multiple candidate tokens are generated in parallel, corresponding to different watermark strength strategies. Naively decoding each candidate sequentially would incur a high computational cost, while direct batch decoding would consume substantial GPU memory when the number of candidates is large. To address this, we adopt the tree attention technique~\cite{miao2024specinfer}, which treats all candidate tokens as child nodes sampled from the same parent distribution. A specially constructed attention mask disables attention between sibling candidates while preserving attention to shared prefix tokens, allowing all candidate branches to be processed in a single forward pass. This approach significantly reduces both latency and memory overhead compared to traditional decoding, enabling efficient integration of the watermark capacity evaluator during generation. Further architectural details can be found in Appendix~\ref{MPM}.

\section{Experiments and Analysis}\label{section4}

\subsection{Experiments Setups}

We use Llama2 and Vicuna-v1.5 models (7B and 13B) to evaluate the generalizability of \myname. Two benchmark datasets are used: MMLU (sociology) to assess factual retention and GSM8K for mathematical reasoning. Unlike open-ended tasks, these datasets provide standardized answers to better reflect watermark impact.
We apply our plug-and-play watermark framework to three base methods: KGW~\cite{kuditipudi2023robust}, Unigram~\cite{zhao2023provable}, and EXP~\cite{kuditipudi2023robust}. 
For quality-enhanced plugins, we compare with the entropy-based strategy~\cite{lee2023wrote}, with detection limited to generation-only access (no prompt/model).
We report AUROC and F1 for detection, and task-specific accuracy for text quality, instead of perplexity, to better capture semantic degradation.
All experiments use 5-shot chain-of-thought prompting with greedy decoding and 200-token limits. Each setup is run five times. Parameters like $\delta$ (for KGW/Unigram) and top-K (for EXP) are tuned. We compare ``hard'' and  ``soft'' watermark settings based on strength. Implementation follows~\cite{pan2024markllm}, using a 32GB V100 GPU.
More implementation details are provided in Appendix~\ref{ES}.

\subsection{Main Results}

Tab~\ref{tab:main} shows that \myname consistently outperforms baselines: with similar detection rates, it achieves higher accuracy, and with similar accuracy, it yields better detection performance.

\xhdr{Reasoning Capabilities}
From the results of the GSM8K dataset, current text watermarking methods still inevitably undermine the emerging capabilities of LLM in logical reasoning.
Compared to the three anchor watermarking methods and the entropy-based baseline method, \myname achieves the optimal trade-off between task accuracy and detection rate.
With few-shot and chain-of-thought prompting, LLM responses to GSM8K tasks include detailed reasoning, providing space for watermark embedding.
Even without watermarks, model performance varies due to differences in architecture and training; for example, Vicuna-v1.5 outperforms Llama2 on reasoning tasks.
However, due to differing probability distributions, achieving similar detection rates across models may require adjusting parameters—e.g., Vicuna requires a larger $\delta$ in vocabulary-based methods than Llama2.
Watermarking methods also vary in effectiveness; EXP shows stronger performance in reasoning compared to vocabulary-based approaches.
Overall, despite differences in model capability and watermark type, all models show consistent accuracy degradation after watermark injection.
As shown in Tab~\ref{tab:main}, \myname significantly improved task accuracy while maintaining the detection rate compared to baselines.
In the 7b and 13b versions of the Llama2 and Vicuna-v1.5 models, \myname achieved the best trade-off between accuracy and detection rate.
The \myname method achieves a detection rate similar to that of the Hard baseline while also attaining accuracy comparable to that of the Soft baseline.
For example, for the Llama2-13b model, \myname improved the detection rate by 28.32\%, 22.31\%, and 41.06\% compared to three baseline watermarking methods, despite the detection rates being similar.

\begin{table*}[t]
  \setlength\tabcolsep{4pt}
  \centering
  \scriptsize
    \caption{Detection rates and task accuracy comparison between baseline methods (KGW, UNI, EXP) with Soft/Hard strength settings and the proposed \myname. 
    Results demonstrate \myname's superior text quality and detection performance through joint optimization.}
    \begin{tabular}{c|c|ccc|ccc|ccc|ccc}
    \toprule
    \multirow{15}[12]{*}{\begin{sideways}GSM8K\end{sideways}} & \multirow{2}[4]{*}{} & \multicolumn{3}{c|}{Llama2-13b} & \multicolumn{3}{c|}{Vicuna-13b-v1.5} & \multicolumn{3}{c|}{Llama2-7b} & \multicolumn{3}{c}{Vicuna-7b-v1.5} \\
\cmidrule{3-14}          &       & Acc.  & AUC   & F1    & Acc.  & AUC   & F1    & Acc.  & AUC   & F1    & Acc.  & AUC   & F1 \\
\cmidrule{2-14}          & No watermark & 0.2760  & -     & -     & 0.3080  & -     & \multicolumn{1}{c}{-} & 0.1500  & -     & \multicolumn{1}{c}{-} & 0.2060  & -     & - \\
\cmidrule{2-14}          & KGW-Hard & 0.1748  & 0.9177  & 0.8552  & 0.1648  & 0.9369  & 0.8728  & 0.0864  & 0.9023  & 0.8307  & 0.0936  & 0.9559  & 0.8894  \\
          & KGW-Soft & 0.2072  & 0.8836  & 0.8140  & 0.2396  & 0.8686  & 0.8020  & 0.1140  & 0.8496  & 0.7818  & 0.1436  & 0.8994  & 0.8253  \\
          & KGW+entropy & 0.2114  & 0.8803  & 0.8073  & 0.2468  & 0.8626  & 0.7978  & 0.1044  & 0.8830  & 0.8154  & 0.1592  & 0.8823  & 0.8049  \\
          & KGW+ours & 0.2243  & 0.9207  & 0.8535  & 0.2680  & 0.9120  & 0.8379  & 0.1105  & 0.9169  & 0.8445  & 0.1572  & 0.9474  & 0.8886  \\
\cmidrule{2-14}          & UNI-Hard & 0.1820  & 0.9003  & 0.8272  & 0.1936  & 0.9421  & 0.8741  & 0.0896  & 0.8939  & 0.8244  & 0.1004  & 0.9440  & 0.8761  \\
          & UNI-Soft & 0.2125  & 0.8515  & 0.7869  & 0.2532  & 0.8481  & 0.7819  & 0.1180  & 0.8013  & 0.7472  & 0.1492  & 0.8660  & 0.7923  \\
          & UNI+entropy & 0.2191  & 0.8694  & 0.8177  & 0.2306  & 0.8537  & 0.7893  & 0.1086  & 0.8447  & 0.7884  & 0.1402  & 0.8812  & 0.7996  \\
          & UNI+ours & 0.2226  & 0.9139  & 0.8460  & 0.2493  & 0.9267  & 0.8522  & 0.1104  & 0.9192  & 0.8517  & 0.1500  & 0.9312  & 0.8582  \\
\cmidrule{2-14}          & EXP-Hard & 0.1744  & 0.9561  & 0.9033  & 0.1820  & 0.9496  & 0.8857  & 0.0816  & 0.9535  & 0.8983  & 0.1388  & 0.9138  & 0.8488  \\
          & EXP-Soft & 0.1968  & 0.9199  & 0.8539  & 0.2148  & 0.8902  & 0.8323  & 0.1076  & 0.9105  & 0.8429  & 0.1576  & 0.8724  & 0.8077  \\
          & EXP+entropy & 0.2036  & 0.9394  & 0.8811  & 0.2352  & 0.9048  & 0.8394  & 0.1060  & 0.9295  & 0.8636  & 0.1600  & 0.8642  & 0.7958  \\
          & EXP+ours & 0.2460  & 0.9462  & 0.8970  & 0.2412  & 0.9445  & 0.8820  & 0.1084  & 0.9640  & 0.9150  & 0.1604  & 0.9032  & 0.8406  \\
    \midrule
    \multirow{13}[8]{*}{\begin{sideways}MMLU\end{sideways}} & No watermark & 0.7114  & -     & -     & 0.6318  & -     & -     & 0.5373  & -     & -     & 0.6219  & -     & - \\
\cmidrule{2-14}          & KGW-Hard & 0.4498  & 0.9310  & 0.8643  & 0.4129  & 0.8783  & 0.8134  & 0.3463  & 0.9368  & 0.8714  & 0.4408  & 0.8796  & 0.8171  \\
          & KGW-Soft & 0.5701  & 0.7881  & 0.7436  & 0.5821  & 0.7419  & 0.7128  & 0.4836  & 0.7814  & 0.7367  & 0.5473  & 0.7116  & 0.6932  \\
          & KGW+entropy & 0.5403  & 0.8835  & 0.8215  & 0.5871  & 0.7159  & 0.6926  & 0.4080  & 0.8678  & 0.8061  & 0.4806  & 0.7895  & 0.7377  \\
          & KGW+ours & 0.5920  & 0.9402  & 0.8718  & 0.6119  & 0.9072  & 0.8401  & 0.5423  & 0.9017  & 0.8304  & 0.5871  & 0.8446  & 0.8118  \\
\cmidrule{2-14}          & UNI-Hard & 0.4119  & 0.8609  & 0.7941  & 0.4876  & 0.8703  & 0.8031  & 0.3622  & 0.8817  & 0.7888  & 0.4308  & 0.8743  & 0.8049  \\
          & UNI-Soft & 0.6189  & 0.6952  & 0.6994  & 0.5970  & 0.7080  & 0.6843  & 0.4527  & 0.7020  & 0.6993  & 0.5025  & 0.7371  & 0.7193  \\
          & UNI+entropy & 0.5871  & 0.7233  & 0.6847  & 0.5771  & 0.7131  & 0.6836  & 0.4378  & 0.8685  & 0.7990  & 0.5224  & 0.8191  & 0.7463  \\
          & UNI+ours & 0.6617  & 0.8627  & 0.7843  & 0.6219  & 0.8829  & 0.8075  & 0.4975  & 0.8985  & 0.8284  & 0.5920  & 0.8902  & 0.8049  \\
\cmidrule{2-14}          & EXP-Hard & 0.5453  & 0.8667  & 0.8138  & 0.4577  & 0.8765  & 0.8279  & 0.4239  & 0.8449  & 0.7984  & 0.4328  & 0.9113  & 0.8486  \\
          & EXP-Soft & 0.5841  & 0.8060  & 0.7594  & 0.5821  & 0.8431  & 0.7712  & 0.4577  & 0.7858  & 0.7605  & 0.5970  & 0.8188  & 0.7592  \\
          & EXP+entropy & 0.5831  & 0.8214  & 0.7762  & 0.5423  & 0.8717  & 0.7933  & 0.4577  & 0.8002  & 0.7645  & 0.5721  & 0.8508  & 0.8088  \\
          & EXP+ours & 0.6269  & 0.9122  & 0.8364  & 0.6020  & 0.9335  & 0.8654  & 0.5124  & 0.8575  & 0.8145  & 0.6070  & 0.9181  & 0.8454  \\
    \bottomrule
    \end{tabular}%
  \label{tab:main}%
\end{table*}%

\xhdr{Knowledge Memory Capabilities}
Results on the MMLU dataset indicate that existing watermarking methods still negatively impact LLMs’ ability to recall memorized knowledge.
Unlike GSM8K, which emphasizes step-by-step reasoning, MMLU focuses on factual recall embedded in the model’s parameters.
We selected the sociology subset, which contains multiple-choice questions testing fixed knowledge points without involving reasoning or computation.
The few-shot, chain-of-thought prompting setup allows the model to analyze the question and options before selecting an answer from A, B, C, or D.
As shown in Tab~\ref{tab:main}, \myname significantly improves accuracy compared to baselines while preserving a high detection rate.
These findings highlight \myname’s advantage in maintaining watermark robustness without compromising the model’s knowledge memory capabilities, offering a more practical and reliable watermarking solution.

\begin{figure*}[t]
    \centering
    \includegraphics[width=0.95\linewidth]{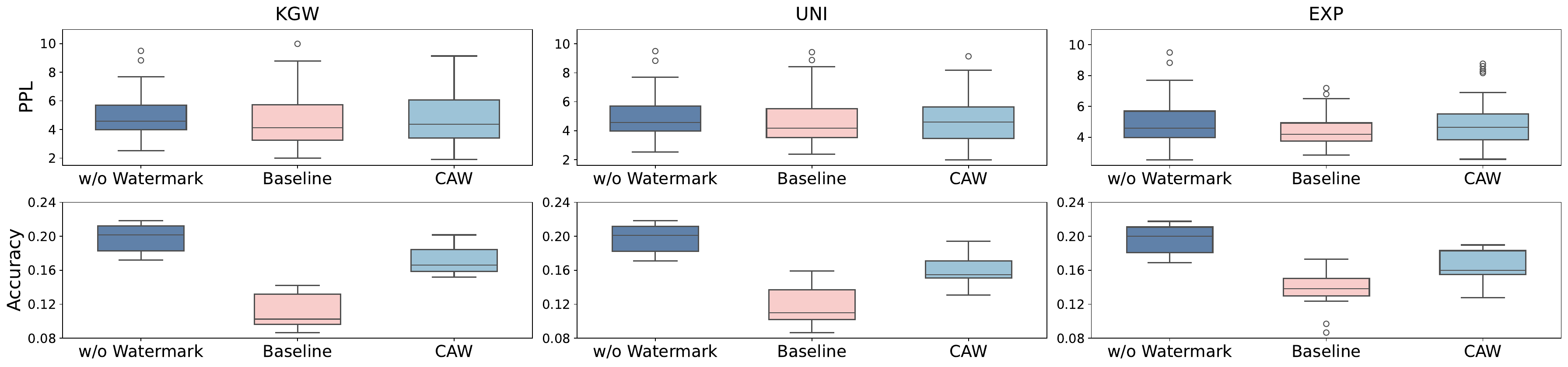} %
    \vspace{-6pt}
    \caption{Box plots of perplexity and accuracy for non-watermarked text, baselines (KGW, UNI, EXP), and \myname.
P-values show no significant perplexity difference but clear accuracy gaps, highlighting perplexity's weakness in reflecting text quality.}
    \label{fig:ppl}
\end{figure*}

\begin{figure*}[t]
    \centering
    \includegraphics[width=0.96\linewidth]{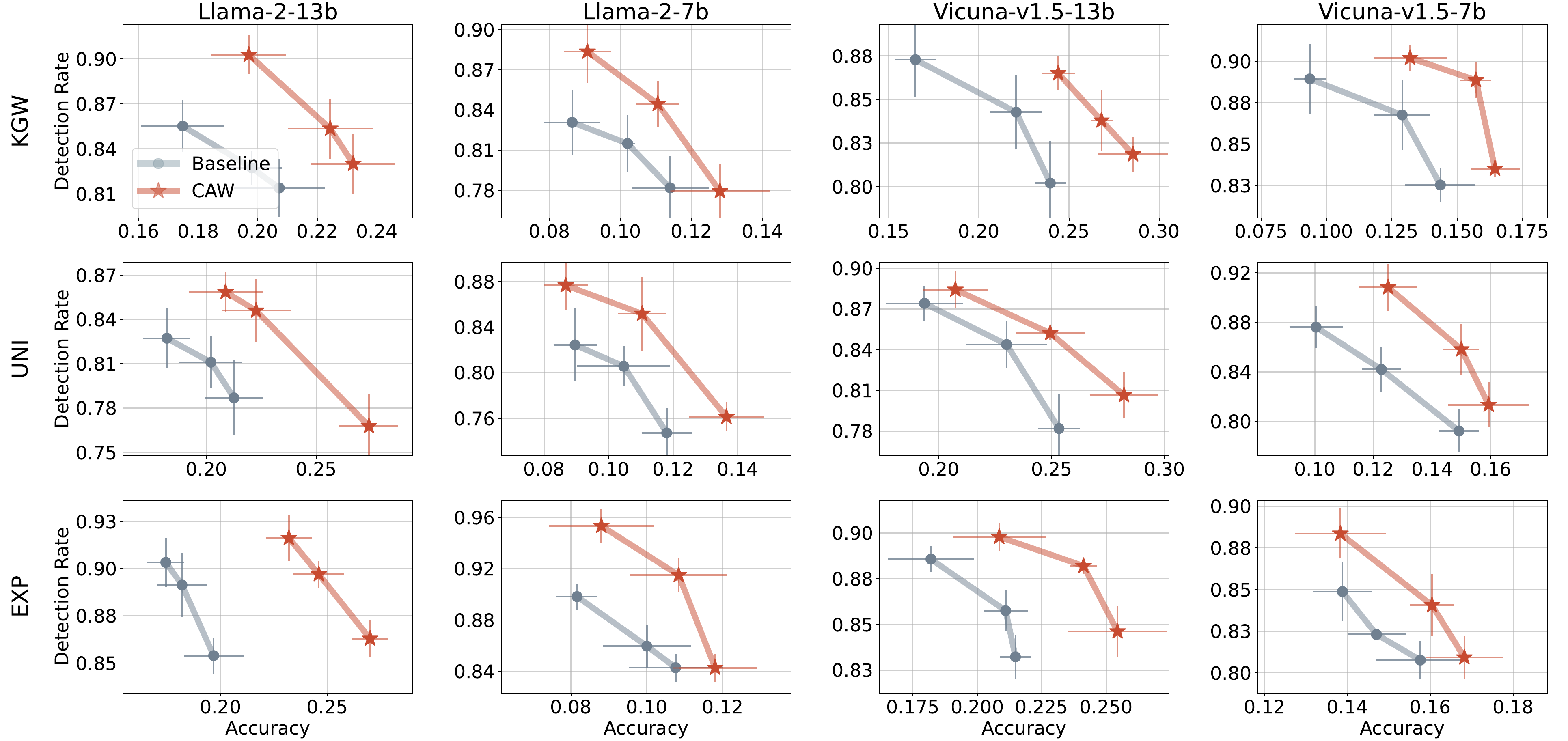} 
    \vspace{-5pt}
    \caption{Pareto front comparisons of \myname and baselines across Llama2 and Vicuna-v1.5 (7B/13B), with Soft, Medium, and Hard watermarking settings. Results show that \myname consistently achieves a better trade-off between detection rate and task accuracy.}
    \label{fig:paleto}
\end{figure*}

\xhdr{Models of different scales} 
To evaluate \myname across model scales, we tested both 7B and 13B LLMs. Although their baseline capabilities differ, they show consistent trends under varying watermark strengths. For example, with the EXP method at around 90\% F1 detection, accuracy on GSM8K dropped by 36.81\% (13B) and 45.60\% (7B) using the baseline, but only 10.97\% (13B) and 27.73\% (7B) with \myname. 
Smaller models are more impacted, likely due to simpler architectures and fewer parameters, making them more sensitive to watermark disruption. 
In contrast, larger models exhibit greater determinism, enhancing capacity evaluation and preserving output quality.

\xhdr{Perplexity Analysis}
In the main experiment, we use task accuracy to assess fine-grained text quality.
Here, we further examine the validity of perplexity. 
Using Vicuna-7B on GSM8K for generation and Llama2-13B for perplexity evaluation, we compared watermark-free, baseline, and \myname outputs. 
As shown in Fig~\ref{fig:ppl}, perplexity distributions showed no significant differences, while task accuracy varied substantially, indicating perplexity fails to reflect real quality differences. 
This highlights the need for task-specific accuracy in watermark evaluation. 
Overall, \myname improves output quality without compromising robustness, making it a practical, plug-and-play enhancement for existing watermarking methods.

\subsection{Trade-off Performance Analysis}
Building on the main results, we further analyzed the detection–quality trade-off between \myname and baseline methods under varying watermark strengths, as illustrated in Fig~\ref{fig:paleto}. For each method, we adjust watermark strength across three settings: soft, mid, and hard. These tiers offer practical insights—soft configurations preserve text quality with moderate detectability, hard ones favor detection at the cost of quality, and mid settings aim for a balance. 
Experiments on both Llama2 and Vicuna-v1.5 models (7B/13B) using GSM8K demonstrate that \myname consistently outperforms its baselines, achieving full Pareto optimality in all cases. 
This dominance holds across all model scales, confirming that \myname offers a generalizable and architecture-agnostic solution for balancing watermark detectability and generation quality.

\begin{figure*}[t]
    \centering
    \includegraphics[width=\linewidth]{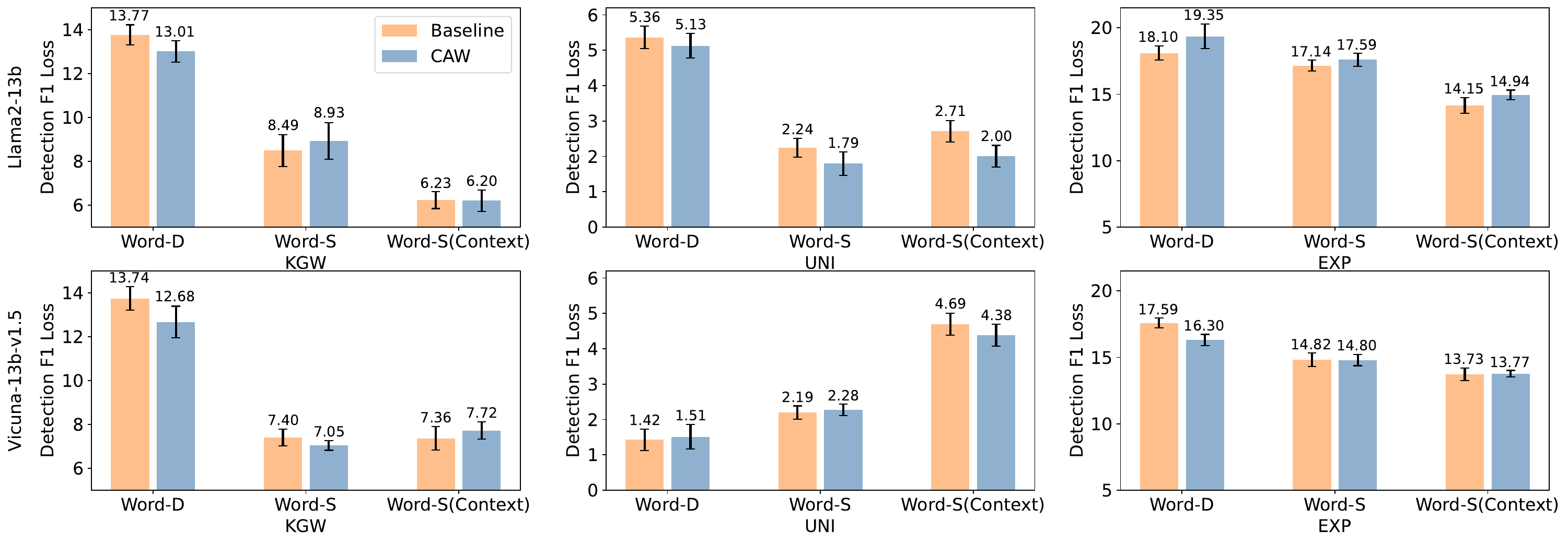} %
    \vspace{-14pt}
    \caption{
    Detection rate loss under three attacks on watermarked Llama2-13B and Vicuna-v1.5-13B, with/without \myname, indicating that \myname preserves original robustness without significant impact.
    }
    \label{fig:robustness}
\end{figure*}

\subsection{Robustness Analysis}

Figure~\ref{fig:robustness} shows the impact of \myname on the attack robustness of the original method.
Here, we conducted experiments using the Llama2-13b and Vicuna-v1.5-13b model on the GSM8K dataset, comparing the attack robustness of three baseline watermarking methods along with the case of \myname.
In this context, the detection rate loss on the y-axis refers to the value by which the detection rate of the watermarking method decreases compared to the original case.
That is, the smaller the detection rate loss, the higher the attack robustness.

We test three attack methods including \textit{Word-S}, \textit{Word-D} and \textit{Word-S-Context}.
\textit{Word-S} method refers to the random synonym replacement of words.
\textit{Word-D} method refers to the random deletion of words, and \textit{Word-S-Context} method refers to replacing synonyms based on contextual relevance using a language model.
For fairness, we maintain the same probability of random replacement or deletion for each method.
For each data point, we conducted five repeated trials to obtain the mean and variance.
Due to the use of a fixed vocabulary partition, the Unigram method itself has a high level of robustness.
In contrast, the other two methods that regenerate the random seed based on prefix tokens exhibit relatively lower robustness.
\textit{Word-D} attack undermines overall robustness most significantly.
In general, \myname has no significant impact on the robustness of various watermarking methods.

\subsection{Latency Analysis}

\begin{wrapfigure}{r}{0.52\linewidth} 
    \centering
    \includegraphics[width=1\linewidth]{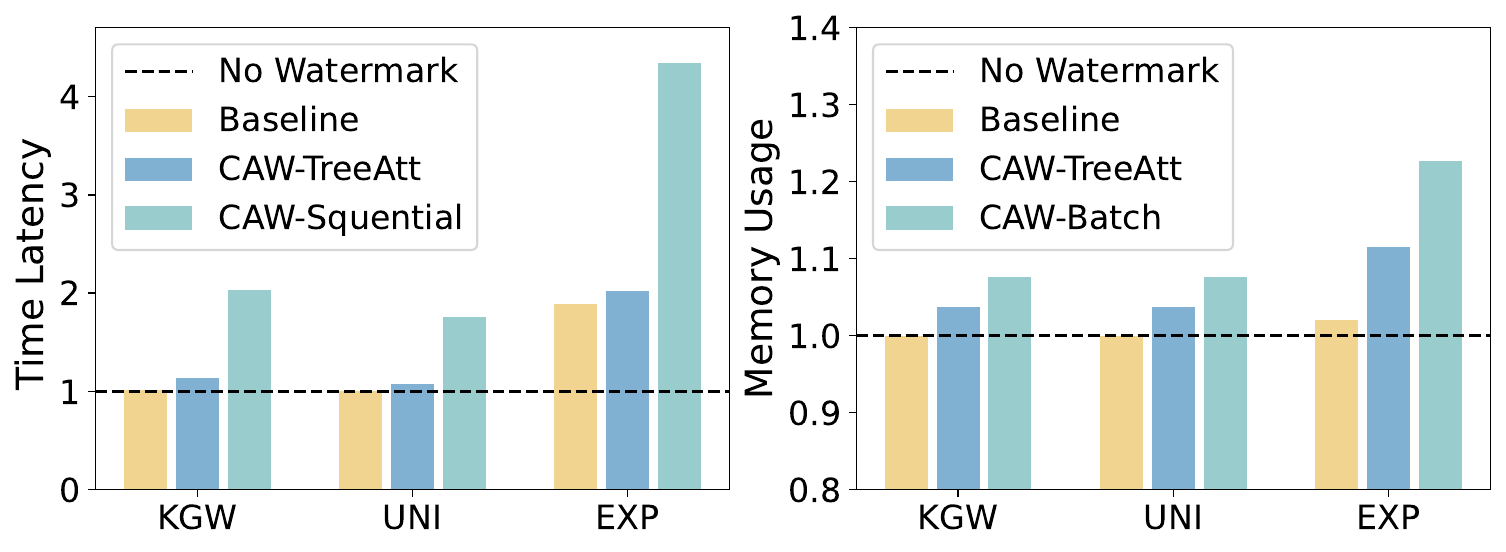} 
    \vspace{-10pt}%
    \caption{    
    Time and memory overhead of different \myname implementations (tree attention, sequential, batch), normalized to the non-watermarked baseline. Tree attention shows clear efficiency advantages.
}
    \label{fig:efficiency}
\end{wrapfigure}

As watermarking algorithms are embedded in LLM inference, minimizing their impact on speed is critical for practical use.
We evaluate the time and memory overhead of \myname under identical model, prompt, and generation configurations, averaging results over 10 runs.
The baseline (set to 1) is the model without watermarking.
Figure~\ref{fig:efficiency} reports relative time and memory usage during inference.
Compared to the three baseline methods, \myname introduces less than 15\% extra latency due to its multi-branch pre-generation mechanism.
Memory overhead remains within 5\% for KGW and UNI, and around 10\% for EXP due to more candidate branches.
Since \myname does not alter the detection process, only generation overhead is analyzed.
As a comparison, we also provide the sequential and batch decoding implementations of \myname, demonstrating the spatiotemporal efficiency advantages of the multi-branch pre-generation mechanism.

\section{Conclusion} \label{section5}
Watermarking has become a significant mechanism for discerning AIGC from human-generated content.
In this paper, we propose \myname, a plug-and-play and model-agnostic watermark capacity-aware watermarking framework, which can be seamlessly integrated with various existing watermarking techniques to enhance generation quality.
Unlike existing watermarking methods that modify the sampling process of LLMs with heuristic rules, 
an adaptive watermarking embedding strategy is employed in \myname to adjust the watermark strength, minimizing potential degradation in content quality.
Specifically, \myname incorporates an online watermarking capacity evaluator to model the watermark capacity at different token positions by analyzing the contextual generation states. 
Moreover, \myname adopts a multi-branch pre-generation mechanism to prevent the decrease in efficiency caused by the use of contextual information during generation.
The effectiveness of our method has been verified through extensive experiments on datasets spanning multiple domains, showcasing superior performance in terms of both detection rate and generation quality compared to various baselines.
Overall, \myname ensures that the watermark remains effective without substantially compromising task-specific performance, enhancing the text generation quality while maintaining the watermark detection effectiveness. 
In the future, we hope to combine our method with more diverse watermarking frameworks, such as multi-bit watermarking and sentence-level watermarking.


\bibliographystyle{unsrtnat}
\bibliography{Bibliography}

\newpage

\appendix

\section{Further Discussion on Related Work} \label{RL}
\xhdr{Existing Efforts on Enhancements of Text Quality}
Currently, some research efforts are also attempting to enhance the output text quality during the watermarking process\cite{hu2023unbiased,lee2023wrote,lu2024entropy, wouters2024optimizing, fu2024watermarking,chen2024watme}.
Some studies have explored unbiased watermarking methods to maintain text quality.
For example, \citet{hu2023unbiased} explore unbiased watermark framework by applying certain reweighting functions to the probability distribution.
Such prompt-dependent detection methods face practical limitations, as original prompts are rarely available, especially for malicious AI-generated text where only the final output is disseminated.
Meanwhile, requirement for the original model also leads to significant computational overhead.
Therefore, despite having good theoretical properties, their practical effectiveness in real-world scenarios is limited.

Other research in this area uses entropy as a criterion for assessing the importance of tokens, enhancing the quality of the watermark text\cite{lee2023wrote,lu2024entropy, liu2024adaptive}.
\citet{lee2023wrote} improved the output quality and detection effectiveness of watermarks by removing low-entropy segments during the generation and detection of watermarks.
\citet{lu2024entropy} propose an entropy-based watermark detection method that gives higher-entropy tokens higher weights during the detection phase. 
\citet{liu2024adaptive} introduce an adaptive watermarking strategy that includes a framework for watermark token identification, which effectively identifies tokens with high entropy distribution.
However, these entropy-based works sometimes require the use of prompts and generation models to reproduce token logits during the detection phase.
In a realistic detection phase, considering time efficiency and availability, this hypothesis of possessing a prompt and generation model is stringent and difficult to achieve.
Furthermore, the entropy of a single isolated token presents challenges related to insufficient information and localized informational bias, lacking contextual semantic awareness. 
This limitation may result in the misidentification of irreplaceable tokens.

Several other works utilize semantic-aware vocabulary partitioning methods to enhance watermarking effectiveness.
\citet{fu2024watermarking} propose a watermarking method for conditional text generation tasks, by considering the semantic similarity between token embeddings while partitioning vocabulary. 
\citet{chen2024watme} leverages linguistic prior knowledge regarding the inherent lexical redundancy of LLM vocabulary to prevent the unavailability of suitable tokens caused by vocabulary partition.
However, these methods can only be applied to vocabulary partition-based watermarking and cannot be used for sampling-based watermarking.
They also overlook that even with embedding similarity, replacing critical words can substantially impair quality, while token replaceability is highly context-dependent.

In conclusion, most current watermarking methods use heuristic rules to modify the token sampling process of language models, inevitably leading to the suboptimal selection of semantically important tokens. 
While numerous studies have explored various approaches to balance watermark detection and the quality of output text, opportunities for further improvement remain.
Therefore, we aim to fully consider the differentiated capacity for watermarking information within text content and make more effective use of the internal states during the inference of large models, jointly optimizing the security of the watermark and the generated text quality.

\section{Problem Formulation} \label{PF}
Here, we formulate the white-box watermarking problem from a holistic perspective.
Note that in our setup, the input prompt and the generative model are not accessible for detection, considering real-world conditions.
The watermarking framework usually consists of two components, the watermark encoding process where the hidden information is embedded within the output of model $\mathcal{M}$, and the watermark detection process where a hypothesis test is applied to the output sequence $\textbf{\textit{y}}$ to identify its source.

Consider a generative language model $\mathcal{M} \in \mathcal{V}^* \rightarrow \Delta(\mathcal{V})$ with a vocabulary $\mathcal{V}$, and a prompt sequence $\textbf{\textit{x}} \in \mathcal{V}^*$.
The model provider shares with the watermark detector a random key $\textbf{\textit{k}} \in \mathcal{K}^*$.
The text generation process with watermarking is denoted as:
\begin{equation}
\textbf{\textit{y}} = Generate_{\text{WM}}(\mathcal{M}, \textbf{\textit{x}}, \textbf{\textit{k}}),
\mathcal{V}^* \times \mathcal{K}^* \rightarrow \mathcal{V}^*
\end{equation}
where $Generate_{\text{WM}}$ represents the generation function with watermarking, and $\textbf{\textit{y}} \in \mathcal{V}^*$ is the output text sequence embedded with secret information.
When model $\mathcal{M}$ calls $Generate_{\text{WM}}$ to perform the generation process, it first calculates the probability distribution $p(t_i) \in \Delta(\mathcal{V})$ at position $i$ independently.
Then, the distribution $p(t_i)$ is reweighted, or the subsequent sampling strategy is modified.
A certain decoding function that varies according to different watermarking methods $f: \mathcal{K} \times  \Delta(\mathcal{V}) \rightarrow \mathcal{V}$ is employed, to map the key element $k_i$ and the distribution $p(t_i)$ to a certain next token.

The detection process is fundamentally a hypothesis testing procedure, with a null hypothesis $H_0$: \textit{The text sequence is generated with no knowledge of $Generate_{\text{WM}}$}.
Then a certain detection procedure, for example, a z-statistic is performed to determine whether $\textbf{\textit{y}}^{\prime} \in \mathcal{V}^*$ has been watermarked:
\begin{equation}
z = Detect(\textbf{\textit{y}}^{\prime}, \textbf{\textit{k}}),
\mathcal{V}^* \times \mathcal{K}^* \rightarrow \mathbb{R}
\end{equation}
If the z-score $z$ exceeds the predetermined threshold, it is considered that $\textbf{\textit{y}}^{\prime}$ carries a watermark, leading to the rejection of the null hypothesis $H_0$.
Given practical considerations, it is stipulated that the model $\mathcal{M}$ and prompt $\textbf{\textit{x}}$ are excluded from the detection process.
This exclusion arises from the potential inability to obtain the corresponding prompt during detection and the considerable time consumption associated with using the original model to regenerate probability distributions.

In general, various watermarking methods employ distinct watermark generation functions \( Generate_{\text{WM}}(\mathcal{M}, \textbf{\textit{x}}, \textbf{\textit{k}}) \) and detection processes \( Detect(\textbf{\textit{y}}, \textbf{\textit{k}}) \). 
Yet, the overall architecture typically adheres to the framework outlined above.

\section{Details of Watermark Capacity Evaluator} \label{WCE}
\begin{figure}
    \centering
    \includegraphics[width=0.5\linewidth]{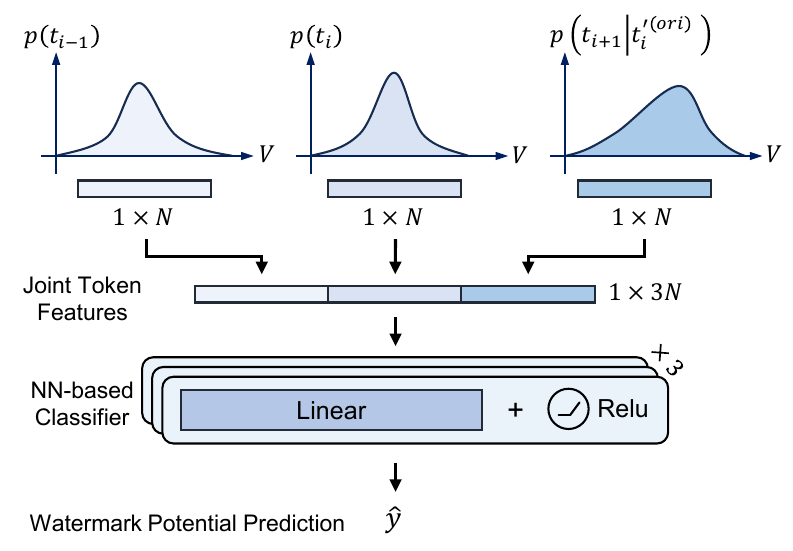} %
    \vspace{-10pt}
    \caption{The structure of the \Cname, which can assess the impact of embedding messages at different token positions.
    It processes the joint probability distribution of three consecutive tokens (current token position $i$ with its contextual neighbors $i-1$ and $i+1$) through a deep neural network-based evaluator to predict the normalized watermark capacity score ($\hat{C_i} \in [0, 1]$) for each token position.   
}
    \label{fig:classifierStructure}
\end{figure}

Let us denote the evaluator as $\mathcal{E}$, this process can be formulated as follows:
\begin{equation}
\label{eq:evaluator}
\hat{C_i} = \mathcal{E}\left( \left[ p(t_{i+k}) \right]_{k=-N^-}^{N^+}; \Theta^{*} \right),
\end{equation}
Here, $\hat{C_i}$ in $(0,1)$ represents the predicted watermark capacity of token $t_i$, and $\Theta^{*}$ represents the well-trained parameter of the classifier.
The parameters $-N^-$ and $-N^+$ regulate the extent of contextual information considered in the input for \(\mathcal{E}\).
The specific architecture of \(\mathcal{E}\) is not restricted.
In practice, we have chosen to implement a three-layer fully connected network.
To evaluate watermark capacity, we establish a task that trains the evaluator and assesses the performance of both our proposed evaluator and other algorithms in modeling watermark capacity. 
This task leverages original human responses from the dataset as reference answers, and utilizes advanced language models (such as GPT-4o) to automatically identify key textual segments that significantly impact output quality, which are then annotated as the ground truth for watermark capacity measurement. 
Utilizing this task, we can train our proposed watermark capacity evaluator.
For simplicity, we denote $\left[ p(t_{i+k}) \right]_{k=-N^-}^{N^+} $ as $\mathbf{p}(t_i)$.
The parameters of \(\mathcal{E}\), denoted as \(\Theta\), as are optimized based on the training set with a cross-entropy optimization objective.
\begin{equation}
\begin{aligned}
    \Theta^{*}=\arg \min _{\Theta}-\frac{1}{|\mathcal{T}|} \sum_{(\mathbf{p}(t_i), \hat{C_i}) \in \mathcal{T}} \hat{C_i} \log (\mathcal{E}(\mathbf{p}(t_i) ; \Theta)) \\
    +(1-\hat{C_i}) \log (\mathcal{E}(\mathbf{p}(t_i) ; \Theta))) .
\end{aligned}
\end{equation}
Figure~\ref{fig:classifierStructure} illustrates the structure of the proposed watermark capacity evaluator.

\section{Mechanisms of Contextual Generation States-Aware Watermarking} \label{CGSA}
The proposed capacity evaluator enables dynamic estimation of each token's watermark capacity, which directly informs its optimal watermark strength.
This adaptive approach maximizes watermark injection while preserving output accuracy, achieving superior trade-offs between detection reliability and text quality.
Specifically, the capacity-to-strength conversion follows a linear mapping scheme with an empirically determined threshold \(\theta\).
Tokens with predicted capacity \(\hat{C_i} > \theta\) receive full protection through the original probability distribution and sampling strategy, safeguarding semantically critical positions.  
For tokens with \(\hat{C_i} < \theta\), the system applies watermarking with strength scaled by their evaluated capacity.

Now, we will discuss combining the watermark capacity evaluator mentioned in the above section with different watermarking methods when we have \(\hat{C_i} < \theta\) at position $i$.
Considering the original probability distribution as \( p(t_i) \), the original sampling strategy without watermark is:
\begin{equation}
    t_i^{(ori)} = \underset {t \in \mathcal{V}} { \operatorname {arg\,max} } \, p(t_i)
\end{equation}
For sampling-based methods, the watermark strength variable can be the restriction of the $Top\text{-}K$ candidate pool during each sampling, or a coefficient in the watermark sampling function.
Here, we will explain with $Top\text{-}K$ being the watermark strength variable for simplicity.
In \textbf{Step 2}, the framework determines all possible watermark strategies, generating all potential candidates for token \( t_i \).
Sampling is performed only among the candidate pool, which contains the tokens that rank in the top-K by probability.
Let the original top-K be \(K\). As mentioned in \ref{subsection33}, for the token \(t_i\), if we already have the prediction of the evaluator from equation~\ref{eq:evaluator}, we can correspondingly calculate \(K^{\prime} = \beta \frac{\theta - \hat{C_i}}{\theta} K\).
However, in \textbf{Step 2}, since \(\textbf{p}(t_i) = [p(t_{i-1}), p(t_i), p(t_{i+1} \vert t_i^{\prime(ori)})\) has not been obtained yet, it is temporarily impossible to calculate \(\hat{y}\) and \(K^{\prime}\).
Therefore, we proactively generate all potential candidates for the token \(t_i\) according to all possible values of \(K^{\prime}\).
All possible values of \(K^{\prime}\) are \([1, 2, ..., \texttt{round}(\beta K)]\).
Although the watermark capacity value output by the NN-based evaluator is continuous, the corresponding watermark strength variable top-k is discrete.
Consider a specific watermark strength variable top-K as \( K \). The set of tokens ranked in the top \( K \) according to \( p(t_i) \) is denoted as \( \mathcal{V}_K \). The corresponding watermark sampling method is:
\begin{equation}
    t_i^{wm} = \underset {t \in topK} { \operatorname {arg\,max} } \, \mathcal{F}(t_i)^{\frac{1}{p(t_i)}}
\end{equation}
where $\mathcal{F}$ is a random mapping function from $\mathcal{V}$ to $\mathbb{R}$, generated using secret key $\textbf{\textit{k}}$ as a random seed.
Since the values of \( K \) are discrete, it is feasible to list all possible \( K \) values that the watermark capacity evaluator might provide, and subsequently enumerate all feasible sampling strategies.

For reweighting-based watermarking methods, the watermark strength variable is the bias value \( \delta \) added to the original probability distribution.
\begin{equation}  
    \delta^{\prime} = \beta \frac{\theta - \hat{C_i}}{\theta} \delta   
\end{equation}  
Notice that in the series of methods based on reweighting-based methods led by KGW, although the watermark strength variable \(\delta\) is continuous, the division of the red-green list is still discrete.
Therefore, the actual sampling either selects the token with the highest logits from the green table or the token with the highest logits from the red table, making the specific strategy still discrete.

In general, the preliminary investigation reveals that LLM-generated text inherently possesses measurable watermark capacity characteristics.
Experimental evidence confirms that the internal inference states of large models reliably indicate watermark capacity potential, with our proposed evaluator demonstrating superior modeling capabilities.

\section{Explanations on Multi-branch Pre-generation Mechanism} \label{MPM}
We now present a detailed introduction to the proposed multi-branch pre-generation mechanism, which is a specific elaboration on Step 2 and Step 3 in subsection~\ref{subsection32}.
As illustrated in Fig~\ref{fig:method}, this mechanism aims to reduce the negative impact of the watermark capacity evaluator on time efficiency in the overall quality-enhancement framework.
While the contextual generation states-aware watermark capacity evaluator effectively determines token importance and corresponding watermark strengths, its dependency on next-token probability distributions introduces computational latency.
Our solution pre-generates reweighting or sampling strategies across different watermark strength levels, producing candidate tokens for each position $i$.
This approach enables parallel LLM decoding while avoiding sequential processing delays, preserving the base model's generation efficiency.

For each sampling strategy, the framework pre-generate logits for all possible candidate tokens \(\mathbf{t}_i^{\prime} = [t_i^{\prime(ori)}, t_i^{\prime(1)}, ..., t_i^{\prime(P)}]\) in Step 3, and perform tree attention decoding \cite{miao2024specinfer} to obtain all candidate distributions \(p(t_{i+1})\).
Here, \(P\) is the total number of sampling strategies, which corresponds to the size of the watermark strength variable \(K\).
Notice that different strategies may correspond to the same \(t_i^{\prime}\), and the number of \(t_i^{\prime}\) candidates must be equal to or less than \(K\).

Here, we elaborate on the details of tree attention.
Most of the widely used LLM architectures use transformers as the backbone, and attention is the core of each transformer layer\cite{vaswani2017attention}.
Currently, most generative language models adopt a decoder-only architecture that utilizes only the transformer decoder part, which has been shown to be highly effective for text generation tasks\cite{wang2022language}.
In decoder-only architecture, the attention calculation is causal. When generating each new token, the model relies only on the previously generated tokens.
To ensure that the model does not use non-causal information when calculating attention, an attention mask strategy is employed, limiting the visibility of the self-attention mechanism.
Let $A$ be the matrix of attention scores, and the causal attention mask is as follows:
\begin{equation} \label{eq:mask}
    \text{mask}(A)_{jk} = 
    \begin{cases}
        A_{jk}, & \text{if } j \geq k \\
        -\infty, & \text{if } j < k
    \end{cases}
    \qquad 0 \leq j \leq i.
\end{equation}
\( A_{jk} \) represents the attention value between token \( t_i \) and \( t_j \), and \(\text{mask}(A)_{jk}\) represents the result of masking the attention value between \( t_i \) and \( t_j \).
In other words, the attention scores for all prefix tokens are retained normally, while the attention scores for subsequent tokens are set to \( -\infty \), indicating that the attention output of the \( j \)-th token is independent of the subsequent tokens.
In Step 3, we need to make the next inference for \( P + 1 \) candidates of \( \textit{t}_i^{\prime} \) simultaneously.
If we calculate \( p(t_{i+1} \vert t_i^{\prime(ori)}) \) for each candidate \( t_i^{\prime(j)} \) sequentially, as shown in Fig~\ref{fig:tree-attn}, it would result in \( P + 1 \) calculations, consuming a substantial amount of computation time. 
For methods where \(P\) can be relatively large, directly treating the \( P + 1 \) token sequences as a batch and allowing the LLM to perform batch decoding would consume a significant amount of GPU memory.

Therefore, we draw on the tree attention decoding method \cite{miao2024specinfer}.
Note that all the candidates \(\textit{t}_i^{\prime}\) are sampled from the same distribution \(p(t_i)\), and they can be viewed as \(P + 1\) child nodes of a parent node.
For each child node \(t_i^{\prime(j)}\), the only meaningful prefix essentially consists of \(t_{(i-1)}\) and the tokens preceding it; the other candidates for \(t_i\) that are at the same level do not have a causal relationship with it.
Therefore, in the attention mask, it is sufficient to set the attention scores of each candidate with respect to the other candidates to \(-\infty\).

Figure~\ref{fig:tree-attn} shows the details of the tree attention decoding for \(\textit{t}_i^{\prime}\).
Let us be at the \(i\)-th step, where the identified prefix tokens are \([t_0, t_1, ..., t_{i-1}]\). This includes the tokens within the prompt and the tokens that have already been generated.
Let all possible candidate tokens be \(\textit{t}_i^{\prime}\), that is, \(\textit{t}_i^{\prime} = [t_i^{\prime(ori)}, t_i^{\prime(1)}, ..., t_i^{\prime(P)}]\). Therefore, for all confirmed prefix tokens \([t_0, t_1, ..., t_{i-1}]\), the attention mask remains unchanged and is still as shown in \ref{eq:mask}.
For the candidate \(\textit{t}_i^{\prime}\), we have:
\begin{equation}
    \text{mask}(A)_{(jk)} = 
    \begin{cases}
        A_{(jk)}, & j \geq k \\
        -\infty, & j < k \text{ or } \psi(j, k)
    \end{cases}
    \quad 0 \leq j \leq i + P,\\
\end{equation}
where $\psi(j, k) \equiv (i \leq j \leq i + P) \text{ and } (i \leq k \leq i + P)$.
Based on the lower triangular attention mask matrix of the basic causal model, \(\text{mask}(A)_{(jk)}\) between \(t_i\) and \(t_j\) that satisfies \(\psi(j, k)\) is also set to \(-\infty\).
The candidates \(t_i\) are in a parallel relationship, and only one of them will be determined as the final result. Therefore, each of them only computes attention with the prefix tokens that form the sequence.
In this way, the operation can be completed in one computation, avoiding the time overhead caused by repeated calculations.
In addition, this approach saves GPU memory resources compared to directly using a batch decoding strategy.

In Step 4 and Step 5, we select the corresponding watermark strength based on the calculated watermark capacity, which means finalizing the watermarking strategy.
For different models, the variables controlling the watermark strength vary. In vocabulary partition-based methods like KGW, the bias value $\delta$ added to the logits of green list tokens represents the watermark strength. In the EXP method, $top-K$ represents the watermark strength.
Once the watermarking strategy is determined, the model can then decide on the selection of token \( t_i \), while also obtaining \( p(t_{i+1} \vert t_i) \).
We use the pre-generated \(p(t_{i+1} \vert t_i)\) to proceed to the next decoding iteration so it will not cause additional latency.

\section{Experiments Setups} \label{ES}
\xhdr{Models}
We use two large language models, Llama2 \cite{touvron2023llama} and Vicuna-v1.5 \cite{chiang2023vicuna}.
The results in this article are obtained using the 7B and 13B versions respectively.
By conducting experiments on models with different scales and training methods, we aim to demonstrate the versatility of \myname.

\xhdr{Baselines}
Our watermark plugin framework can be directly integrated with existing watermarking algorithms.
Therefore, we selected three established watermarking algorithms as base watermark algorithms to be integrated with different watermark capacity control algorithms.
The KGW algorithm\cite{kuditipudi2023robust} randomly splits the vocabulary into a green list and a red list, and biases tokens in the green list to facilitate detection. 
The Unigram algorithm\cite{zhao2023provable} builds on the KGW algorithm by using a fixed hash key during watermarking to enhance attack robustness.
The EXP algorithm\cite{kuditipudi2023robust} employs a random hash mapping function and biases the selection towards tokens with higher hash values during watermark detection.
We chose the entropy-based method\cite{lee2023wrote} as the baseline for the watermark capacity control algorithm.
Before performing watermark sampling using the base method, it first calculates the entropy corresponding to the probability distribution of the current token. 
When the entropy exceeds a predetermined threshold, watermarking is no longer applied to that token.
Unlike the original method, we take into account more realistic scenarios. Considering efficiency and feasibility, we do not allow the model and prompt to participate during the detection phase.
In other words, token entropy can only affect the selection during the generation phase, and have no influence on the detection phase.

\xhdr{Datasets}
We selected two representative datasets with standard answers: the MMLU \cite{hendrycks2020measuring} multiple-choice question dataset and the GSM8K \cite{cobbe2021training} mathematical application problem dataset.
The MMLU dataset, from which we selected the subject of sociology, is primarily used to assess the extent to which watermarks disrupt the model's memory of the knowledge itself.
The GSM8K dataset is mainly designed to evaluate the model's symbolic computation and reasoning capabilities.
These two datasets examine the fundamental capabilities of large models from different perspectives, reflecting the extent to which watermarks can disrupt the capabilities of large language models.
Unlike the high-degree-of-freedom datasets such as text completion and story generation commonly used in existing works\cite{raffel2020exploring}, the dataset we have selected can more comprehensively reflect the extent to which watermarks disrupt the accuracy of text.

\xhdr{Evaluation Metrics}
The evaluation of watermarks focuses on two metrics: detection rate and text quality.
Consistent with previous work, we use the Area Under the Receiver Operating Characteristic curve (AUROC) and F1 score, two well-established metrics for binary classifiers, as indicators for watermark detection rate.
As for text quality, based on previous discussions, we are no longer using perplexity as a metric. Instead, we directly use accuracy corresponding to the task.
Compared to perplexity, metrics corresponding to the task reflect the accuracy of the model's answers, better demonstrating the impact of the watermark on the model's knowledge retention and logical reasoning abilities.

\xhdr{Other Details}
For both datasets, a 5-shot prompting method is used in the experiments.
At the same time, the chain of thought technique \cite{wei2022chain} is also used in prompts, allowing the model to fully demonstrate its thought process when responding.
In this way, the model's responses contain information that should be protected, which would affect the overall correctness of the answer if compromised. 
At the same time, it also includes tokens of relatively lower importance that can withstand a certain degree of replacement.
Greedy search is adopted as the sampling method for all models and datasets.
The maximum number of generated tokens for all tasks is set to 200, ensuring the complete output of the answers.
For each data point in Tab~\ref{tab:main} and Fig~\ref{fig:paleto}, we conducted five repeated experiments.
The key parameters of different watermarking methods have been adaptively adjusted for different models and datasets.
In the vocabulary-based method, \(\gamma\) (the proportion of green tokens in the whole vocabulary) is consistently set to 0.25, with only \(\delta\) as the key parameter to control the watermark strength.   
In the EXP method, the hash function remains unchanged, and only top-K serves as the key parameter.
Both the baseline method and \myname can adjust the watermark strength parameter.
We refer to implementations with a stronger watermark strength, that is, a higher detection rate and lower accuracy, as hard baselines, while those with a lower watermark strength are referred to as soft baselines.
The baseline implementation is based on the open-source project from \cite{pan2024markllm}.
As for hardware, the main experiments are conducted on a 32GB NVIDIA V100 GPU.

\section{Limitation}
The datasets used in this study are primarily concentrated on typical and widely used benchmarks. Although these datasets are representative, our findings may not fully generalize to emerging new domains with unique knowledge characteristics. Future work should aim to expand the scope of validation to improve the applicability and generalizability of the method. Additionally, our current approach focuses on specific watermarking frameworks. In the future, we hope to integrate our method with more diverse watermarking techniques, such as multi-bit watermarking and sentence-level watermarking, to further enhance its robustness and flexibility.


\end{document}